# On local structures of gravity-free space and time

Jian-Miin Liu*
Department of Physics, Nanjing University
Nanjing, The People's Republic of China
*On leave. E-mail address: liu@mail.davis.uri.edu

Besides two fundamental postulates, (i) the principle of relativity and (ii) the constancy of the one-way speed of light in all inertial frames of reference, the special theory of relativity uses the assumption about the Euclidean structure of gravity-free space and the homogeneity of gravity-free time in the usual inertial coordinate system. Introducing the so-called primed inertial coordinate system, in addition to the usual inertial coordinate system, for each inertial frame of reference, we assume the flat structures of gravity-free space and time in the primed inertial coordinate system and, hence, their generalized Finslerian structures in the usual inertial coordinate system. We further modify the special theory of relativity by combining the alternative assumption with the two postulates (i) and (ii). The modified special relativity theory involves two versions of the light speed, infinite speed c' in the primed inertial coordinate system and finite speed c in the usual inertial coordinate system. It involves the c'-type Galilean transformation between any two primed inertial coordinate systems and the localized Lorentz transformation between any two usual inertial coordinate systems. The physical principle in the modified special relativity theory is: the c'-type Galilean invariance in the primed inertial coordinate system plus the transformation from the primed to the usual inertial coordinate systems. It combines with the quantum mechanics theory to found a convergent and invariant quantum field theory. We produce corrections to the Maxwellian velocity and velocity rate distributions for free particles. The detection of velocity and velocity rate distributions for free particles can serve as the tests to investigate the local structures of gravity-free space and time.

## I. INTRODUCTION

The current Lorentz-invariant field theory, whether classical or quantum, has been suffering from the divergence difficulties for a long time. Indeed, the infinite self-energy of an electron in quantum electrodynamics was known as early as 1929 [1], while that in classical electrodynamics was known earlier. The origins of these difficulties lay deep within the conceptual foundations of the theory. Two foundation stones of the current quantum field theory are the special relativity theory and the quantum mechanics theory. Since it is the case that both classical and quantum field theory are plagued by the divergence difficulties, the direction to get to the roots of these difficulties seems to be in the special relativity theory.

According to the special relativity theory, the energy of a particle is not an invariant, and the transformation properties of the total energy of many-particle systems are completely not clear under the Lorentz transformation because simultaneity at distant space points in a given inertial frame of reference is no longer simultaneous in any different inertial frame of reference. Also, relativistic time appears to be incompatible with the canonical notion of evolution of many-particle systems. These and other, such as the Lorentz non-invariance of box volume, cause much difficulties in constructing the Lorentz-invariant statistical mechanics and the Lorentz-invariant thermodynamics of many-particle systems in the framework of the special relativity theory [2].

Although the special theory of relativity has been enormously successful in explaining various physical phenomena and has been tested to very high degree of precision, the above difficulties in quantum field theory and Lorentz-invariant statistical mechanics indicate that it is not a ultimate theory. A kind of modification is needed for it. Probably just because of this, the question "Is the special theory of relativity, for reasons unspecified and unknown, only an approximate symmetry of nature?" was raised [3], and the neutrino and Kaon tests were proposed to investigate possible violations of the local Lorentz invariance [3,4].

In this paper, we are concerned with a new notion about the local structures of gravity-free space and time, based on which we propose a modified special relativity theory. The paper consists of twelve sections. In the next section, we analyze experimental facts for indication how to modify the special theory of relativity. The experimental indication is quite explicit. With this indication, we go back to the special relativity theory in Section III and show that it actually uses the assumption about the flat structures of gravity-free space and time in the usual inertial coordinate system besides two fundamental postulates on the principle of relativity and the speed of light. We are going to make our assumption about the local structures of gravity-free space and time, in connection with generalized Finsler geometry. But, since Finsler geometry and its generalization are not familiar to many people, we make an instruction to Finsler geometry and generalized Finsler geometry first, in Section IV, and our assumption later, in Section V. Section VI addresses the modified special relativity theory that is formed by combining the alternative assumption with the two fundamental postulates. Section VII is devoted to the physical principle in the modified special relativity theory. Then, in Sections VIII and IX, we sketch the applications of this principle to reform of mechanics and field theory. We also sketch how the modified special relativity theory and the quantum mechanics theory together found a convergent and invariant quantum field theory. Readers who are interested in detailed derivations may refer to Ref.[24]. There, the knowledge of the Cartan connection in the generalized Finsler geometry is necessary. In Section X, we discuss the velocity-space in the modified special relativity theory. This discussion eventually leads to corrections to the Maxwellian velocity and velocity rate distributions of free particles, which are shown in Section XI. Finally, in Section XII, we make some concluding remarks.

II. EXPERIMENTAL INDICATION

Because of technological limitations, in the earlier experiments testing the constancy or isotropy of the light speed, light was propagated in a closed path. The favorite conclusions from these experiments are explicitly for the constancy of the speed of the round-trip propagating light, not for that of the one-way speed of light. As a result of the technological development, many experiments have been done in the manner that light propagates in a one-way. Turner and Hill [5] and Champeney et al [6], both research groups did experiments using the Mossbauer effect. They placed a $Co^{57}$ source near the rim of a standard centrifuge with an iron absorber near the axis of rotation. The Mossbauer effect was to look for any velocity dependence of the frequency of the 14.4 KeV γ-rays as seen by the $Fe^{57}$ in the absorber. Their experiments established limits of $\Delta c/c < 2 \times 10^{-10}$ for the anisotropy in the one-way speed of light. Riis and his colleagues [7] compared the frequency of a two-photon transition in a fast atomic beam to that of a stationary absorber while the direction of the fast beam is rotated relative to the fixed stars. They found the upper limit $\Delta c/c < 3.5 \times 10^{-9}$ firstly and $\Delta c/c < 2 \times 10^{-11}$ later for the anisotropy. The experiment of Krisher et al [8] was made by use of highly stable hydrogen-maser frequency standards (clocks) separated by over 21 km and connected by a ultrastable fiber optics link. They compared the phases of the standards and produced the limits from the experimental data, $\Delta c/c < 2 \times 10^{-7}$ and $\Delta c/c < 2 \times 10^{-8}$ for linear and quadratic dependencies, respectively, on the velocity of the Earth with respect to the cosmic microwave background.

All experimental tests of the constancy of the one-way light speed can be also interpreted as the tests of the local Lorentz invariance. Nevertheless, since local Lorentz non-invariance implies a departure from the Einstein time dilation and singles out a preferred inertial frame of reference, the experiments done by McGowan et al [9], Bailey et al [10], Kaivola et al [11], Prestage et al [12], and Krisher et al [8] can be accounted immediate testing the local Lorentz invariance. Bailey et al, Kaivola et al, and McGowan et al verified the Einstein time dilation to an accuracy of $1 \times 10^{-3}$, $4 \times 10^{-5}$ and $2.3 \times 10^{-6}$ respectively. Experiments of Prestage et al and Krisher et al are sensitive to the effects of motion of their experimental apparatus relative to a preferred inertial frame of reference and sensitive to the form of time dilation coefficient of the used hydrogen-maser clocks. The explicitly null results were yielded in these two experiments for breakdown of the local Lorentz invariance.

Experiments clearly support the existence of the constancy of the one-way light speed, the Einstein velocity addition law and the local Lorentz invariance [5-13]. The experimental indication is: any modification of the special relativity theory must keep the constancy of the one-way speed of light and the local Lorentz invariance.

III. THE SPECIAL RELATIVITY THEORY

Einstein published his special theory of relativity in 1905 [14]. He derived the Lorentz transformation between any two usual inertial coordinate systems, which is the kinematical background for the physical principle of the Lorentz invariance. The two fundamental postulates stated by Einstein as the basis for his theory are (i) the principle of relativity and (ii) the constancy of the one-way speed of light in all inertial frames of reference. Besides these two fundamental postulates, the special theory of relativity also uses another assumption. This other assumption concerns the Euclidean structure of gravity-free space and the homogeneity of gravity-free time in the usual inertial coordinate system $\{x^r,t\}$, $r=1,2,3$, $x^1=x$, $x^2=y$, $x^3=z$,

$$dX^2=\delta_{rs}dx^r dx^s, \quad r,s=1,2,3, \tag{1a}$$
$$dT^2=dt^2, \tag{1b}$$

everywhere and every time.

Postulates (i) and (ii) and the assumption Eqs.(1) together yield the Lorentz transformation between any two usual inertial coordinate systems [14-18]. Indeed, though the assumption Eqs.(1) was not explicitly articulated, evidently having been considered self-evident, Einstein said in 1907: "Since the propagation velocity of light in empty space is c with respect to both reference systems, the two equations, $x_1^2+y_1^2+z_1^2-c^2t_1^2=0$ and $x_2^2+y_2^2+z_2^2-c^2t_2^2=0$, must be equivalent." [17]. Leaving aside a discussion of whether postulate (i) implies the linearity of transformation between any two usual inertial coordinate systems and the reciprocity of relative velocities between any two usual inertial coordinate systems, we know that the two equivalent equations, the linearity of transformation and the reciprocity of relative velocities lead to the Lorentz transformation.

Some physicists explicitly articulated the assumption Eqs.(1) in their works on the topic. Pauli wrote: "This also implies the validity of Euclidean geometry and the homogeneous nature of space and time." [16], Fock said: "The logical foundation of these methods is, in principle, the hypothesis that Euclidean geometry is applicable to real physical space together with further assumptions, viz. that rigid bodies exist and that light travels in straight lines." [18].

Introducing the four-dimensional usual inertial coordinate system $\{x^\gamma\}$, $\gamma=1,2,3,4$, $x^4=ict$, and the Minkowskian structure of gravity-free space-time in this coordinate system,

$$d\Sigma^2=\delta_{\alpha\beta}dx^\alpha dx^\beta, \quad \alpha,\beta=1,2,3,4, \tag{2}$$

Minkowski [19] showed in 1909 that the Lorentz transformation is just a rotation in four-dimensional space-time. He also showed how to use the four-dimensional tensor analysis for writing invariant physical laws under the Lorentz transformation. The Minkowskian structure Eq.(2) is a four-dimensional version of the assumption Eqs.(1).

IV. GENERALIZED FINSLER GEOMETRY

Finsler geometry is a kind of generalization of Riemann geometry [20-22]. It was first suggested by Riemann as early as 1854, and studied systematically by Finsler in 1918. Since then, the most significant work from the viewpoint of modern differential geometry has been done by Cartan, Rund and others.

In Finsler geometry, distance ds between two neighboring points $x^k$ and $x^k+dx^k$, $k=1,2,---,n$ is defined by a scale function

$$ds=F(x^1,x^2,---,x^n;dx^1,dx^2,---dx^n)$$

or simply

$$ds=F(x^k,dx^k), \quad k=1,2,---,n, \tag{3}$$

which depends on directional variables $dx^k$ as well as coordinate variables $x^k$. Apart from several routine conditions like smoothness, the main constraint imposed on this scale function is that it is positively homogeneous of degree one in $dx^k$,

$$F(x^k,\lambda dx^k)=\lambda F(x^k,dx^k) \quad \text{for } \lambda>0. \tag{4}$$

Introducing a set of equations

$$g_{ij}(x^k,dx^k)=\partial^2 F^2(x^k,dx^k)/2\partial dx^i \partial dx^j, \quad i,j=1,2,---,n, \tag{5}$$

we can represent Finsler geometry in terms of

$$ds^2=g_{ij}(x^k,dx^k)dx^i dx^j, \tag{6}$$

where $g_{ij}(x^k,dx^k)$ is called the Finslerian metric tensor induced from scale function $F(x^k,dx^k)$. The Finslerian metric tensor is symmetric in its subscripts and all its components are positively homogeneous of degree zero in $dx^k$,

$$g_{ij}(x^k,dx^k)=g_{ji}(x^k,dx^k), \tag{7a}$$
$$g_{ij}(x^k,\lambda dx^k)=g_{ij}(x^k,dx^k) \quad \text{for } \lambda>0. \tag{7b}$$

We can define the so-called generalized Finsler geometry by omitting Eq.(3) as a definition of ds and instead taking Eq.(6) as a definition of ds$^2$, where the given metric tensor $g_{ij}(x^k,dx^k)$ satisfies Eqs.(7) and

$$\det[g_{ij}(x^k,dx^k)] \neq 0,$$

and

$$g_{ij}(x^k,dx^k)dx^i dx^j > 0, \quad i,j=1,2,---,n,$$

for non-zero vector $dx^i$.

The generalized Finsler geometry is so-named because a Finsler geometry must be a generalized Finsler geometry but the inverse statement is not valid, in other words, a generalized Finsler geometry is not necessarily a Finsler geometry [21].

A generalized Finsler geometry with a given metric tensor $g_{ij}(x^k,dx^k)$ is also a Finsler geometry when and only when we can find a scale function $F(x^k,dx^k)$ by solving the set of equations, Eqs.(5), such that this function is not only positive if not all $dx^k$ are simultaneously zero but also positively homogeneous of degree one in $dx^k$. In this case, $F^2(x^k,dx^k)$ is positively homogeneous of degree two in $dx^k$. The following equation thus holds due to Euler's theorem on homogeneous functions,

$$2F^2(x^k,dx^k) = dx^i[\partial F^2(x^k,dx^k)/\partial dx^i]. \tag{8}$$

Iterating Eq.(8), we find

$$4F^2(x^k,dx^k) = dx^j \delta_{ij}[\partial F^2(x^k,dx^k)/\partial dx^i]$$
$$+ dx^j dx^i[\partial^2 F^2(x^k,dx^k)/\partial dx^j \partial dx^i]. \tag{9}$$

It then follows from use of Eq.(8) that

$$F^2(x^k,dx^k) = dx^j dx^i[\partial^2 F^2(x^k,dx^k)/2\partial dx^j \partial dx^i]. \tag{10}$$

Eq.(10) combines with Eqs.(5) and (6) to yield Eq.(3).

Like Finsler geometry, generalized Finsler geometry can be endowed with the Cartan connection. Relying on that, we can define absolute defferentials and covariant partial derivatives of vectors and tensors.

V. A NEW ASSUMPTION ON LOCAL STRUCTURES OF SPACE AND TIME

Conceptually, the principle of relativity means that there exists a class of equivalent inertial frames of reference, any one of which moves with a non-zero constant velocity relative to any other. Einstein wrote: "in a given inertial frame of reference the coordinates mean the results of certain measurements with rigid (motionless) rods, a clock at rest relative to the inertial frame of reference defines a local time, and the local time at all points of space, indicated by synchronized clocks and taken together, give the time of this inertial frame of reference."[15]. As defined by Einstein, each of the inertial frames of reference is supplied with motionless, rigid unit rods of equal length and motionless, synchronized clocks of equal running rate. Then, in each inertial frame of reference, an observer can employ his own motionless-rigid rods and motionless-synchronized clocks in the so-called "motionless-rigid rod and motionless-synchronized clock" measurement method to measure space and time intervals. By using this "motionless-rigid rod and motionless-synchronized clock" measurement method, the observer in each inertial frame of reference can set up his own usual inertial coordinate system. Postulate (ii) means that the speed of light is measured to be the same constant c in every such usual inertial coordinate system.

The "motionless-rigid rod and motionless-synchronized clock" measurement method is not the only one that each inertial frame of reference has. We imagine, for each inertial frame of reference, other measurement methods that are different from the "motionless-rigid rod and motionless-synchronized clock" measurement method. By taking these other measurement methods, an observer in each inertial frame of reference can set up other inertial coordinate systems, just as well as he can set up his usual inertial coordinate system by taking the "motionless-rigid rod and motionless-synchronized clock" measurement method. We call these other inertial coordinate systems the unusual inertial coordinate systems.

The conventional belief in flatness of gravity-free space and time is natural. But question is, in which inertial coordinate system the gravity-free space and time directly display their flatness. The special theory of relativity recognizes the usual inertial coordinate system, as shown in Eqs.(1). Making a different choice, we take one of the unusual inertial coordinate systems, say $\{x'^r,t'\}$, r=1,2,3, the primed inertial coordinate system. We assume that gravity-free space and time possess the flat metric structures in the primed inertial coordinate system, and hence, the following generalized Finslerian structures in the usual inertial coordinate system,

$$dX^2 = \delta_{rs} dx'^r dx'^s = g_{rs}(y) dx^r dx^s, \quad r,s=1,2,3, \tag{11a}$$

$$dT^2 = dt'^2 = g(y) dt^2, \tag{11b}$$

$$g_{rs}(y) = K^2(y)\delta_{rs}, \tag{11c}$$

$$g(y) = (1 - y^2/c^2), \tag{11d}$$

$$K(y) = \frac{c}{2y} \ln \frac{c+y}{c-y} (1 - y^2/c^2)^{1/2}, \tag{11e}$$

where $y = (y^s y^s)^{1/2}$, $y^s = dx^s/dt$, $s=1,2,3$.

Two metric tensors $g_{rs}(y)$ and $g(y)$ depend only on directional variables $y^s$, $s=1,2,3$, and become flat when and only when y approaches zero.

## VI. THE MODIFIED SPECIAL RELATIVITY THEORY

We modify the special theory of relativity by combining the alternative assumption Eqs.(11), instead of the assumption Eqs.(1), with the two postulates (i) and (ii).

If we define a new type of velocity, $y'^s = dx'^s/dt'$, $s=1,2,3$, in the primed inertial coordinate system and keep the well-defined usual (Newtonian) velocity in the usual inertial coordinate system, we find from the assumption Eqs.(11)

$$y'^s = [\frac{c}{2y} \ln \frac{c+y}{c-y}] y^s, \quad s=1,2,3, \tag{12}$$

and

$$y' = \frac{c}{2} \ln \frac{c+y}{c-y}, \tag{13}$$

where $y' = (y'^s y'^s)^{1/2}$, $s=1,2,3$. It is understood that two different measurement methods can be applied to a motion when the motion is observed in an inertial frame of reference, one being the "motionless-rigid rod and motionless-synchronized clock" measurement method, the other one being associated with the primed inertial coordinate system. As a result, two different velocities are obtained, primed velocity $y'^s$ (of the new type) and usual velocity $y^s$. These two velocities are related by Eqs.(12) and (13). Velocities $y'^s$ and $y^s$ are two versions of the motion obtained via two different measurement methods used in the inertial frame of reference. The Galilean addition among primed velocities links up with the Einstein addition among usual velocities [23]. This statement can be easily seen in the one-dimensional case that

$$y'_2 = y'_1 - u' = (c/2)\ln[(c+y_1)/(c-y_1)] - (c/2)\ln[(c+u)/(c-u)]$$

and

$$y'_2 = (c/2)\ln[(c+y_2)/(c-y_2)]$$

imply

$$y_2 = (y_1 - u)/(1 - y_1 u/c^2).$$

In Eq.(13), as y goes to c, we get an infinite primed speed,

$$c' = \lim_{y \to c} \frac{c}{2} \ln \frac{c+y}{c-y}. \tag{14}$$

Speed c' is invariant in the primed inertial coordinate systems simply because of the invariance of speed c in the usual inertial coordinate systems. Speed c' is really a new version of the light speed, its version in the primed inertial coordinate systems.

Let IFR1 and IFR2 be two inertial frames of reference, where IFR2 moves with a non-zero constant velocity relative to IFR1. IFR1 and IFR2 can use their own "motionless-rigid rod and motionless-synchronized clock" measurement methods and set up their own usual inertial coordinate systems $\{x^r_m, t_m\}$, $m=1,2$. They can also set up their own primed inertial coordinate systems, $\{x'^r_m, t'_m\}$, $m=1,2$. Since the propagation velocity of light is c' in both $\{x'^r_1, t'_1\}$ and $\{x'^r_2, t'_2\}$, we have two equivalent equations,

$$\delta_{rs} dx'^r_1 dx'^s_1 - c'^2 (dt'_1)^2 = 0, \tag{15a}$$

$$\delta_{rs} dx'^r_2 dx'^s_2 - c'^2 (dt'_2)^2 = 0. \tag{15b}$$

Using Eqs.(11) with y=c, we have further two equivalent equations,

$$\delta_{rs}dx^r_1 dx^s_1 - c^2(dt_1)^2 = 0, \qquad (16a)$$
$$\delta_{rs}dx^r_2 dx^s_2 - c^2(dt_2)^2 = 0, \qquad (16b)$$

because $c^2 K^2(c) = c'^2 g(c)$, where $K(c) = \lim_{y \to c} K(y)$, $g(c) = \lim_{y \to c} g(y)$.

Two equivalent equations Eqs.(15), the linearity of transformation between two $\{x'^r_m, t'_m\}$, the reciprocity of relative velocities between two $\{x'^r_m, t'_m\}$, and the flat structures of gravity-free space and time in the primed inertial coordinate systems will lead to the c'-type Galilean transformation between two primed inertial coordinate systems $\{x'^r_m, t'_m\}$, m=1,2, under which speed c' is invariant. Two equivalent equations Eqs.(16), the linearity of transformation between two $\{x^r_m, t_m\}$, and the reciprocity of relative velocities between two $\{x^r_m, t_m\}$ will lead to the localized Lorentz transformation between two usual inertial coordinate systems $\{x^r_m, t_m\}$, m=1,2, where the space and time differentials take places of the space and time variables in the Lorentz transformation.

In the modified special relativity theory, the c'-type Galilean transformation stands between any two primed inertial coordinate systems, while the localized Lorentz transformation between two corresponding usual inertial coordinate systems. Substituting the assumption Eqs.(11) for the assumption Eqs.(1) does not spoil the localized Lorentz transformation between any two usual inertial coordinate systems and the local Lorentz invariance in the usual inertial coordinate systems.

## VII. PHYSICAL PRINCIPLE IN THE MODIFIED SPECIAL RELATIVITY THEORY

As said, experiments clearly support the existence of the constancy of the one-way light speed c, the Einstein law governing additions of usual velocities, and the local Lorentz invariance. These facts not only verify the postulates (i) and (ii) but also tell us that it is the "motionless-rigid rod and motionless-synchronized clock" measurement method which we use in our experiments. All our experimental data are collected and expressed in the usual inertial coordinate system.

The physical principle in the special theory of relativity is the Lorentz invariance: In the usual inertial coordinate system, all physical laws keep their forms with respect to the Lorentz transformation. In the modified special relativity theory, we may also chose the usual inertial coordinate system for theoretical purposes. The physical principle is the local Lorentz invariance: In the usual inertial coordinate system, all physical laws keep their forms with respect to the localized Lorentz transformation. This local Lorentz invariance must be implemented in dealing with the non-flat structures Eqs.(11). The modified special relativity theory, however, offers another choice: the primed inertial coordinate system. Taking this choice, we have a new physical principle: The c'-type Galilean invariance in the primed inertial coordinate system plus the transformation from the primed inertial coordinate system to the usual inertial coordinate system. In the primed inertial coordinate system, we write all physical laws in the c'-type Galilean-invariant form, we do calculations in the c'-type Galilean-invariant manner, and we finally transform all results from the primed inertial coordinate system to the usual inertial coordinate system and compare them to experimental facts in the usual inertial coordinate system.

The transformation from the primed to the usual inertial coordinate systems is an important part of the new physical principle. If two axes $x'^r$ and $x^r$, where r runs 1,2,3, are set to have the same direction and the same origin, the transformation is

$$dx'^r = K(y) dx^r, \quad r=1,2,3, \qquad (17a)$$
$$dt' = (1 - y^2/c^2)^{1/2} dt. \qquad (17b)$$

This transformation links the Galilean transformation between any two primed inertial coordinate systems up with the localized Lorentz transformation between two corresponding usual inertial coordinate systems [24].

## VIII. MECHANICS AND FIELD THEORY AND QUANTUM FIELD THEORY

The new physical principle has been applied to reform of mechanics and field theory [24]. The validity of relativistic mechanics in the usual inertial coordinate system remains, while field theory is freshened. In the usual inertial coordinate system, interaction-free field equations contain non-flat structures $g_{rs}(y)$ and $g(y)$. These equations still have the de Broglie wave solution. In the usual inertial coordinate system, the canonically conjugate variables to field variables $\varphi_k$ (vector or spinor) are

$$\pi^k = \partial L / \partial(\partial \varphi_k / \partial t'). \qquad (18)$$

They form a contravariant vector or spinor of the same rank as the original field vector or spinor. In the usual inertial coordinate system, the Noether theorem for energy and momentum has the shape of

$$\frac{d}{dt'}[\frac{1}{ic}\int T^4_j d\mathbf{x'}]=0, \qquad (19a)$$

where

$$T^4_j = L g^4_j(z) - [\partial L/\partial(\partial \varphi_k/\partial x^4)][\partial \varphi_k/\partial x^j] \qquad (19b)$$

and $d\mathbf{x'}=dx'^1 dx'^2 dx'^3$. The conserved field energy and momentum, $\frac{1}{ic}\int T^4_j d\mathbf{x'}$, j=1,2,3,4, form a vector as well as $T^4_j$ because $d\mathbf{x'}$ is invariant. The gauging procedure adopted in the current field theory to make a field system locally gauge-invariant with respect to a certain gauge group is still effective in the primed inertial coordinate system. As we act the transformation from the primed to the usual inertial coordinate systems on this procedure, we can find its version in the usual inertial coordinate system. Moreover, it is the most remarkable feature of the freshened field theory that the concept of particle size has its own room. Particle size defined in the primed inertial coordinate system is an invariant quantity. It can be quite involved in our invariant calculations, though the concept of particle size in the primed inertial coordinate system is different from that we tried but failed to introduce in the current Lorentz-invariant field theory. We discuss it in detail elsewhere.

In the primed inertial coordinate system, any field system can be quantized by use of the canonical quantization method. As instantaneity is a covariant concept, the equal-time commutation or anti-commutation relations are reasonable. As the canonically conjugate variables are contravariant to the original field variables, the commutation or anti-commutation relations are also of tensor equations. As the light speed c' is infinite, the essentially instantaneous quantum connection is acceptable. As the primed time differential dt' is invariant in all inertial frames of reference, the time-ordered product in the perturbation expansion of S-matrix is definite.

In regard to question: How is a quantized field system transformed under the transformation from the primed to the usual inertial coordinate systems, there is a transformation law which has been established. The transformation law says: Any quantized field system will undergo a unitary transformation as it is transformed from the primed inertial coordinate system to the usual inertial coordinate systems [24].

IX. RESOLUTION OF THE DIVERGENCE PROBLEM

The divergence difficulties in the current quantum field theory have been already ascribed to the model of point particle. But, in the framework of the special relativity theory, all attempts to assign a finite size to a particle failed. According to the special relativity theory, particle size is not a covariant concept. It has been unknown how to carry out the Lorentz-invariant calculations based on such a size. The Lorentz invariance recognizes the light speed c constituting a limitation for transport of matter or energy and transmission of information or causal connection. The law of causality rejects any instantaneous processes between two events at distinct space points. It is very hard to explain how a sized particle as a whole is set in motion when a force acts on it at its edge. It is very hard to explain how the quantum connection specified by the quantum mechanics theory can be instantaneous. Also, as pointed out by Poincare [25], the field energy and momentum of a sized electron, in the framework of the special relativity theory, do not have correct transformation properties, unlike its mechanical energy and momentum, though they are finite. That will result in different values when we compute some quantities, f.g. the total mass of the sized electron.

All of these alter in the modified special relativity theory. In the primed inertial coordinate system, particle size is a covariant concept; the light speed constituting a limitation for the matter or energy transport and the information or causal connection transmission is c', infinite; the field energy and momentum of a sized particle form a vector, as well as its mechanical energy and momentum. In the primed inertial coordinate system, all particles properly exhibit their size. As a matter of course, we have a convergent and c'-type Galilean-invariant quantum field theory for any quantized field system in the primed inertial coordinate system. When we perform the transformation from the primed to the usual inertial coordinate systems, the unitary transformation does not change the eigenvalue spectrums of operators of this quantized field system, the expectation values of its observations, and its operator and state vector equations. We hence have a convergent and invariant quantum field theory for the quantized field system in the usual inertial coordinate system. Since it is the localized Lorentz transformation that stands between any two usual inertial coordinate systems, we indeed have a convergent and locally Lorentz-invariant quantum field theory for the quantized field system in the usual inertial coordinate system.

## X. VELOCITY-SPACE IN THE MODIFIED SPECIAL RELATIVITY THEORY

From Eqs.(11) we have

$$Y^2 = [\frac{c}{2y} \ln \frac{c+y}{c-y}]^2 \delta_{rs} y^r y^s \text{ or } Y = \frac{c}{2} \ln \frac{c+y}{c-y}, \tag{20}$$

where $Y = dX/dT$ is the velocity-length. The velocity-space embodied in Eq.(20) is a velocity-space that Fock was the first to define by

$$dY^2 = H_{rs}(y) dy^r dy^s, \ r,s=1,2,3, \tag{21a}$$
$$H_{rs}(y) = c^2 \delta^{rs}/(c^2-y^2) + c^2 y^r y^s/(c^2-y^2)^2, \ -c \leq y^s \leq c, \tag{21b}$$

in the usual velocity-coordinates $\{y^r\}$, $r=1,2,3$. In fact, with standard calculation techniques in Riemann geometry, one can derive Eq.(20) from Eqs.(21). This velocity-space is characterized by the finite velocity boundary c and the Einstein velocity addition law [18,23]. Non-flat generalized Finslerian structures of gravity-free space and time in the usual inertial coordinate system match this Riemannian structure of the velocity-space in the usual velocity-coordinates $\{y^r\}$, $r=1,2,3$.

We can re-defined this velocity-space in the primed velocity-coordinates $\{y'^r\}$, $r=1,2,3$, by

$$dY^2 = \delta_{pq} dy'^p dy'^q, \ p,q=1,2,3, \tag{22a}$$

with

$$dy'^p = A^p{}_r(y) dy^r \tag{22b}$$

and

$$A^p{}_r(y) = \gamma \delta^{pr} + \gamma(\gamma-1) y^p y^r / y^2, \tag{22c}$$

where $y^2 = y^s y^s$, $s=1,2,3$, $\gamma = (1-y^2/c^2)^{-1/2}$ because

$$\delta_{pq} A^p{}_r(y) A^q{}_s(y) = H_{rs}(y). \tag{23}$$

## XI. VELOCITY-DISTRIBUTION TESTS

Eq.(22a) convinces us of validity of the Maxwellian formulas of velocity and velocity rate distributions for free particles in the primed velocity-coordinates $\{y'^r\}$, $r=1,2,3$,

$$P(y'^1,y'^2,y'^3) dy'^1 dy'^2 dy'^3 = N(\frac{m}{2\pi K_B T})^{3/2} \exp[-\frac{m}{2 K_B T}(y')^2] dy'^1 dy'^2 dy'^3, \tag{24a}$$

$$P(y') dy' = 4\pi N (\frac{m}{2\pi K_B T})^{3/2} (y')^2 \exp[-\frac{m}{2 K_B T}(y')^2] dy', \tag{24b}$$

where $y' = (y'^s y'^s)^{1/2}$. Using Eqs.(12), (13), (22b) and (22c), we can represent these two formulas and obtain,

$$P(y^1,y^2,y^3) dy^1 dy^2 dy^3 = N \frac{(m/2\pi K_B T)^{3/2}}{(1-y^2/c^2)^2} \exp[-\frac{mc^2}{8 K_B T}(\ln \frac{c+y}{c-y})^2] dy^1 dy^2 dy^3, \tag{25a}$$

$$P(y) dy = \pi c^2 N \frac{(m/2\pi K_B T)^{3/2}}{(1-y^2/c^2)} (\ln \frac{c+y}{c-y})^2 \exp[-\frac{mc^2}{8 K_B T}(\ln \frac{c+y}{c-y})^2] dy. \tag{25b}$$

In the modified special relativity theory, $P(y^1,y^2,y^3)$ and $P(y)$ are the velocity and velocity rate distributions of free particles in the usual velocity-coordinates $\{y^r\}$, $r=1,2,3$. They reduce to the Maxwellian distributions only in the part of y small enough. It is a characteristic that $P(y^1,y^2,y^3)$ and $P(y)$ go to zero as y approaches c as well as y approaches zero. The detection of velocity and velocity rate distributions for free particles in the molecular or atom or electron beam experiments can serve as the tests to investigate the local structures of gravity-free space and time. It might provide an evidence for the modified special relativity theory.

## XII. CONCLUDING REMARKS

There exists an equivalence class of the inertial frames of reference, any one of which moves with a non-zero constant velocity relative to any other. Each of the inertial frames of reference is supplied with motionless, rigid unit rods of equal length and motionless, synchronized clocks of equal running rate. Each inertial frame of reference can set up its own at least two distinct inertial coordinate systems, the usual inertial coordinate system and the primed inertial coordinate system.

The special theory of relativity is formed by combining the two postulates (I) and (ii) and the assumption that gravity-free space and time possess the flat structures in the usual inertial coordinate system. The modified special relativity theory is formed by combining the two postulates (I) and (ii) and the alternative assumption that gravity-free space and time possess the flat structures in the primed inertial coordinate system and their generalized Finslerian structures in the usual inertial coordinate system. Both the special relativity theory and the modified special relativity theory involve the local Lorentz invariance in the usual inertial coordinate systems and the constancy of the one-way speed of light. But the modified special relativity theory involves more. It owns a new version of the light speed, infinite primed speed c' in the primed inertial coordinate systems. In turn, the modified special relativity theory has a new physical principle: the c'-type Galilean invariance in the primed inertial coordinate system plus the transformation from the primed inertial coordinate system to the usual inertial coordinate system.

In the framework of the modified special relativity theory, relativistic mechanics is still valid in the usual inertial coordinate system, while field theory is freshened. Especially, the modified special relativity theory and the quantum mechanics theory together found a convergent and invariant quantum field theory.

In the modified special relativity theory, the velocity-space can be defined in two different ways, in the usual velocity-coordinates $\{y^r\}$, r=1,2,3, and in the primed velocity-coordinates $\{y'^r\}$, r=1,2,3. This characteristic is absent in the special relativity theory. This characteristic leads us to corrections to the Maxwellian velocity and velocity rate distributions of free particles. These corrections have experimental means of testing the local structures of gravity-free space and time.

ACKNOWLEDGMENT
The author greatly appreciates the teachings of Prof. Wo-Te Shen. The author thanks Prof. Mark Y.-J. Mott, Dr. Allen E. Baumann and Dr. C. Whitney for helpful suggestions.

REFERENCES
[1] W. Heisenberg and W. Pauli, Z. Phys., 56, 1 (1929); 59, 168 (1930); I. Waller, Z. Phys., 62, 673 (1930); J. R. Oppenheimer, Phys. Rev., 35, 461 (1930); W. Heisenberg, Z. Phys., 90, 209 (1934); P. A. M. Dirac, Pro. Camb. Phil. Soc., 30, 150 (1934)
[2] P. G. Bergmann, Phys. Rev., 84, 1026 (1951); R. Hakim, J. Math. Phys., 8, 1315 (1967); C. K. Yuen, Amer. J. Phys., 38, 246 (1970)
[3] S. Coleman and S. L. Glashow, Phys. Lett., B405, 249 (1997)
[4] S. L. Glashow et al, Phys. Rev. D56, 2433 (1997); T. Hambye, R. B. Mann and U. Sarkar, Phys. Lett., B421, 105 (1998); Phys. Rev., D58, 025003 (1998)
[5] K. C. Turner and H. A. Hill, Phys. Rev., 134, B252 (1964)
[6] D. C. Champeney, G. R. Isaak and A. M. Khan, Phys. Lett., 7, 241 (1963)
[7] E. Riis et al, Phys. Rev. Lett., 60, 81 (1988)
[8] T. P. Krisher et al, Phys. Rev., D45, 731 (1990)
[9] R. W. McGowan et al, Phys. Rev. Lett., 70, 251 (1993)
[10] J. Bailey et al, Nature (London), 268, 301 (1977)
[11] M. Kaivola et al, Phys. Rev. Lett., 54, 255 (1985)
[12] J. D. Prestage et al, Phys. Rev. Lett., 54, 2387(1985)
[13] R. M. Barnett et al, Rev. Mod. Phys., 68, 611(1996), p655
[14] A. Einstein, Ann. Physik, 17, 891 (1905)

[15]     A. Einstein, Autobiographical Notes, in: A. Einstein: Philosopheo-Scientist, ed. P. A. Schipp, 3rd edition, Tudor, New York (1970)
[16]     A. Einstein, H. A. Lorentz, H. Minkowski and H. Weyl, The Principle of Relativity, collected papers with notes by A. Sommerfield, Dover, New York (1952); A. S. Eddington, The Mathematical Theory of Relativity, Cambridge University Press, Cambridge (1952); C. Moller, The Theory of Relativity, Oxford and New York (1952); W. Pauli, Theory of Relativity, Pergamon Press Ltd., New York (1958), trans. G. Field
[17]     A. Einstein, Jarbuch der Radioaktivitat und Elektronik, $\underline{4}$, 411 (1907), reprinted in The Collected Papers of A. Einstein, vol.$\underline{2}$, 252, Princeton University Press, Princeton, NJ (1989)
[18]     V. Fock, The Theory of SpaceTime and Gravitation, Pergamon Press (New York, 1959)
[19]     H. Minkowski, Phys. Z., $\underline{10}$, 104 (1909)
[20]     P. Finsler, Uber Kurven und Flachen in Allgemeinen Raumen, Dissertation, Gottingen 1918, Birkhauser Verlag, Basel (1951); E. Cartan, Les Espaces de Finsler, Actualites 79, Hermann, Paris (1934); H. Rund, The Differential Geometry of Finsler Spaces, Springer-Verlag, Berlin (1959)
[21]     G. S. Asanov, Finsler Geometry, Relativity and Gauge Theories, D. Reidel Publishing Company, Dordrecht (1985)
[22]     Jian-Miin Liu, The generalized Finsler geometry, to be published and e-printed
[23]     Jian-Miin Liu, Galilean Electrodynamics (Massachusetts, USA), $\underline{8}$, 43 (1997); $\underline{9}$, 73 (1998); Velocity-space in the modified special relativity theory, to be published
[24]     Jian-Miin Liu, Modification of special relativity and formulation of convergent and invariant quantum field theory, e-printed in hep-th/9805004 at xxx.lanl.gov (the Archive, Los Alamos, NM, USA), to be published; Modification of special relativity and the divergence problem in quantum field theory, to be published and e-printed
[25]     H. Poincare, Comtes Rendus (Paris), $\underline{40}$, 1504 (1905); La Mechanique Nouvelle, Gauthier-Villars (Paris, 1924)